\documentclass[aps,prl,graphicx,superscriptaddress,showkeys,showpacs,reprint]{revtex4-1}
\usepackage{graphicx,natbib,url}

\begin{document}

\title{Charge Exchange and Energy Loss of Slow Highly Charged Ions in 1\,nm Thick Carbon Nanomembranes} 

\author{Richard A. Wilhelm}
\email[Author to whom correspondence should be addressed. Electronic mail: ]{r.wilhelm@hzdr.de}
\affiliation{Helmholtz-Zentrum Dresden-Rossendorf, Institute of Ion Beam Physics and Materials Research, 01328 Dresden, Germany, EU}
\affiliation{Technische Universit\"at Dresden, 01069 Dresden, Germany, EU}
\author{Elisabeth Gruber}
\affiliation{TU Wien - Vienna University of Technology, Institute of Applied Physics, 1040 Vienna, Austria, EU}
\author{Robert Ritter}
\affiliation{TU Wien - Vienna University of Technology, Institute of Applied Physics, 1040 Vienna, Austria, EU}
\author{Ren\'e Heller}
\affiliation{Helmholtz-Zentrum Dresden-Rossendorf, Institute of Ion Beam Physics and Materials Research, 01328 Dresden, Germany, EU}
\author{Stefan Facsko}
\affiliation{Helmholtz-Zentrum Dresden-Rossendorf, Institute of Ion Beam Physics and Materials Research, 01328 Dresden, Germany, EU}
\author{Friedrich Aumayr}
\affiliation{TU Wien - Vienna University of Technology, Institute of Applied Physics, 1040 Vienna, Austria, EU}

\date{\today}

\begin{abstract}
Experimental charge exchange and energy loss data for the transmission of slow highly charged Xe ions through ultra-thin polymeric carbon membranes are presented. Surprisingly, two distinct exit charge state distributions accompanied by charge exchange dependent energy losses are observed. The energy loss for ions exhibiting large charge loss shows a quadratic dependency on the incident charge state indicating that equilibrium stopping force values do not apply in this case. Additional angle resolved transmission measurements point on a significant contribution of elastic energy loss. The observations show that regimes of different impact parameters can be separated and thus a particle's energy deposition in an ultra-thin solid target may not be described in terms of an averaged energy loss per unit length.
\end{abstract}

\pacs{34.35.+a, 34.50.Bw, 34.70.+e, 68.49.Sf, 68.65.-k, 79.20.Rf}
\keywords{slow highly charged ion, HCI, carbon nanomembranes, CNM, stopping power, ion charge state, charge exchange}

\maketitle 

\begin{figure}
\includegraphics[width=\columnwidth]{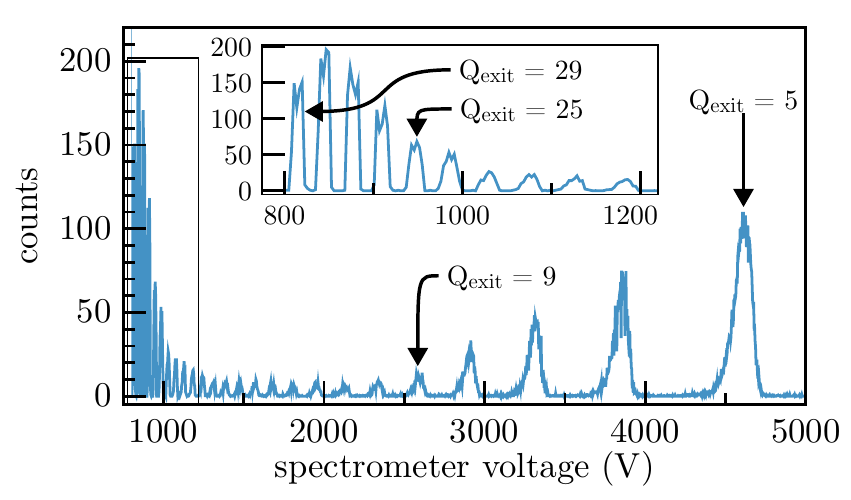} \\ 
\caption{(color online) Spectrum of a 1050\,eV/amu Xe$^{30+}$ beam transmitted through a 1\,nm thick carbon nanomembrane. All charge states below $Q=30$ (but larger than $Q=4$) are visible, whereas two distinct distributions can be observed. The high exit charge state distribution is magnified in the inset.}
\label{fig1}%
\end{figure}

\begin{figure}
\includegraphics[width=\columnwidth]{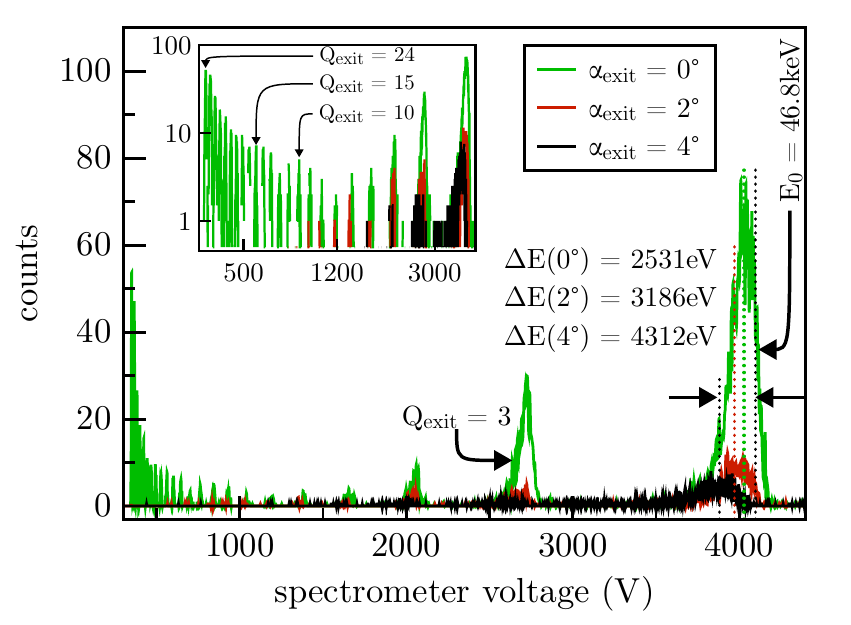} \\ 
\caption{(color online) Transmission spectra of a 46.8\,keV (363\,eV/amu) Xe$^{25+}$ beam for different transmission angles of 0\,$^{\circ}$ (green), 2\,$^{\circ}$ (red) and 4\,$^{\circ}$ (black).  High exit charge states are only observable under 0\,$^{\circ}$ (see double logarithmic inset). Due to the lower beam energy the lowest observable exit charge state is $Q_{exit} = 2$.}
\label{fig2}%
\end{figure}

\begin{figure}
\includegraphics[width=\columnwidth]{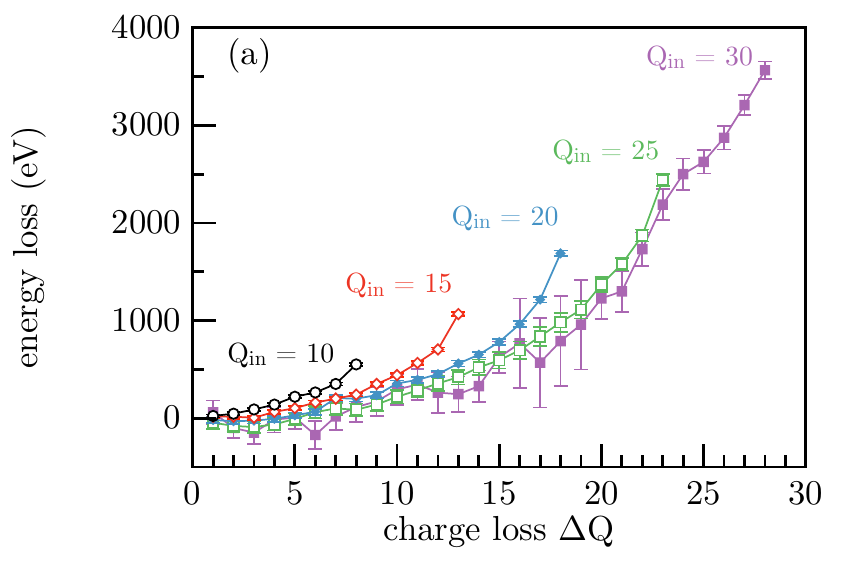} \\ 
\includegraphics[width=\columnwidth]{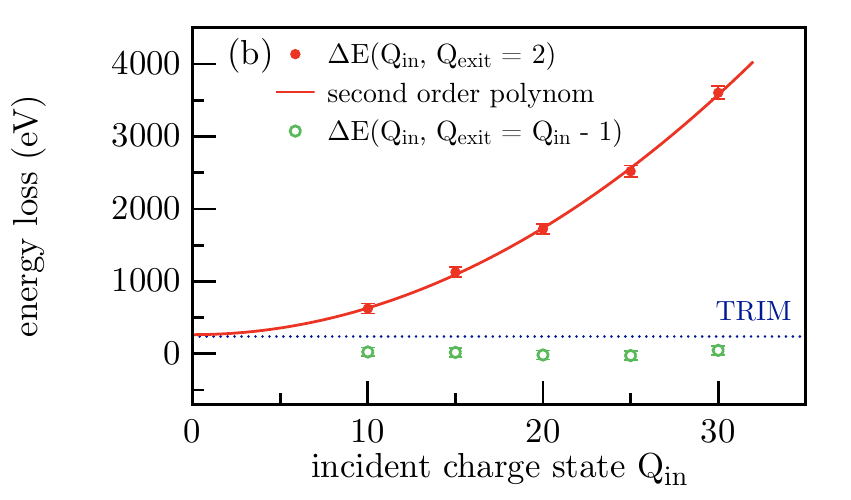} 
\caption{(color online) Energy loss of 40\,keV ions as function of the charge loss $\Delta Q$ (a) and of the incident charge state $Q_{in}$ (b), respectively. In (b) the energy loss is shown only for ions with exit charge state $Q_{exit}=2$ (red) and for ions with a charge loss of $\Delta Q=1$ (green) (maximum and minimum $\Delta Q$). The red curve is a polynomial fit of second order to the obtained data ($\Delta E(Q_{in}) = 3.7$\,eV$\cdot  Q_{in}^2 + 264$\,eV). The dotted line shows the result from a TRIM simulation (see text).}
\label{fig3}%
\end{figure}

\begin{figure}
\includegraphics[width=\columnwidth]{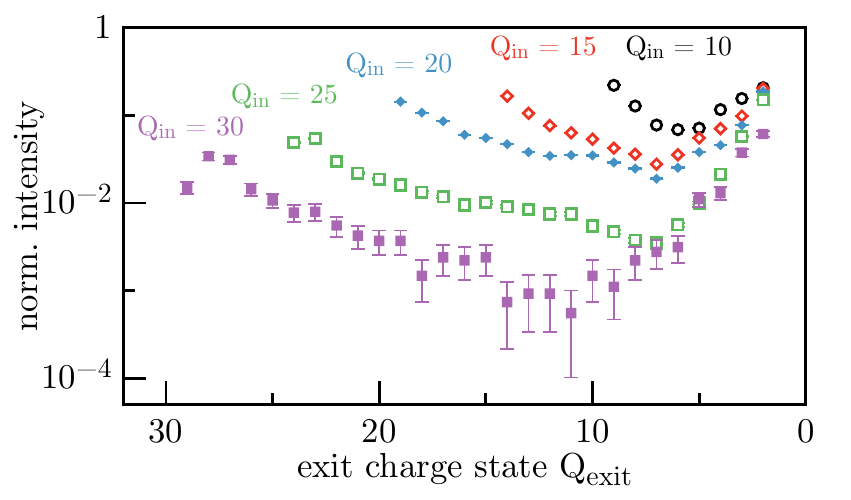} \\ 
\caption{(color online) Normalized intensity of exit charge states for different incident charge states from $Q_{in}=10$ (black) to $Q_{in}=30$ (purple) at $E_{kin}=40$\,keV (310\,eV/amu).}
\label{fig5}%
\end{figure}

\begin{figure}
\includegraphics[width=\columnwidth]{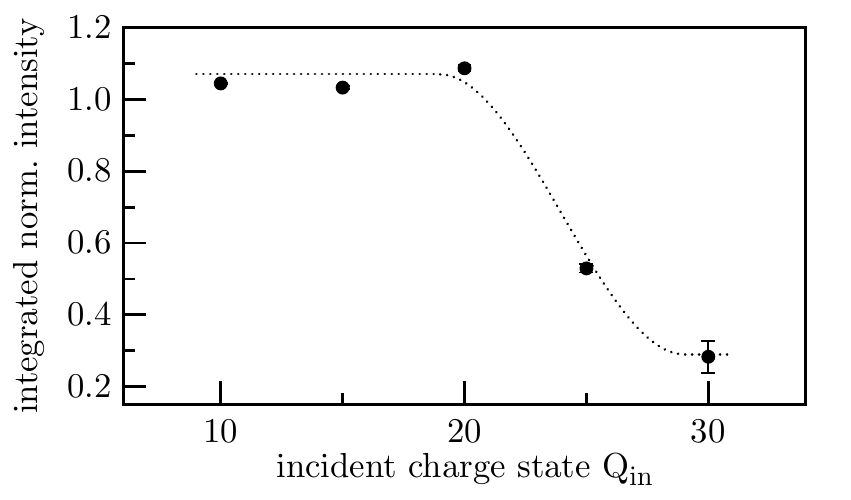} \\ 
\caption{Integrated normalized intensity (see text) as a function of the incident charge state for $E_{kin}=40$\,keV (310\,eV/amu). The dotted line is drawn to guide the eye.}
\label{fig6}%
\end{figure}

Modern approaches in ion and electron irradiation of solids such as nano-structuring of thin films or even structuring of free-standing monolayers such as graphene \cite{GN07,G09,OKB13} or MoS$_2$ \cite{BBK11,KKL13} rely on models for structural and electronic defect formation. Most important for processes during ion-solid interaction is the amount of deposited energy and its dissipation channels \cite{KFM07}. We show that the energy loss and charge exchange of ions in very thin films, such as 2D-materials, show significant differences to solids with reduced thickness. The understanding of these differences is not only of importance for ion beam analysis of 2D-materials but in particular for manipulating and tailoring their properties. \cite{BPK13}. \\
To probe interaction processes in very thin target materials slow highly charged ions (HCI) are ideal tools due to their energy deposition confinement to shallow surface regions. Besides the well known near-surface potential energy deposition \cite{AAE97,AFE11} also an expected increased pre-equilibrium kinetic energy loss (stopping force) \cite{B93} is confined to a few nm at the surface. In the conventional description of both contributions to the stopping force, i.e. nuclear and electronic stopping, the charge state of an ion is identified with its equilibrium charge state by Bohr's stripping criterion \cite{BK82,ZBZ08}. The equilibrium charge state by Bohr is given as $Q_{eq} = Z^{1/3} v/v_0$ and describes the (average) charge state of an ion passing through a solid at a given velocity $v$ ($v_0$: Bohr's velocity, $Z$: nuclear charge of the ion). The charge state $Q$ of slow highly charged ions is much higher than the equilibrium charge state $Q_{eq}$ ($Q_{eq} \ll Q \lesssim Z$). Therefore, the interaction of HCI with surfaces may not be described in terms of an equilibrium charge state dependent stopping force. Furthermore, due to the localization of the energy deposition slow HCI can be used as an efficient tool for surface nano-structuring \cite{HFW08,EHM08,EHA10,EWH12,RWG12,RMB99,PPL05,IWS12,TWT07,BNT05,TTN05,NTN05} and tuning of the electrical properties of materials \cite{PGP07}, as well as a probe for surface energy deposition processes \cite{LPG11,PLS11}. \\
Recently, it has been shown that slow HCI can create pores in 1\,nm thick carbon nanomembranes (CNM) \cite{TBN09,TG12} mainly by deposition of their potential energy \cite{RWS13}. Here we report on measurements of kinetic energy loss and charge loss of slow highly charged Xe ions transmitted through 1\,nm CNM. For carbon foils with larger thicknesses of 5 and 10\,nm Schenkel \textit{et al.} found evidence for a charge state dependent stopping force, whereas the total increase was reported to be small (factor 1.5) \cite{BSS97,SBB97,SHB99} for ions at about 2\,keV/amu. This can be attributed to the fact that the equilibrium charge state is reached within the foil thickness and pre-equilibrium stopping force values may only contribute to a minor extent. In contrast, we observe two \textit{distinct} exit charge state distributions with charge states much higher than the equilibrium charge state and an increase in stopping force with charge state by a factor up to 4, indicating that a 1\,nm carbon layer is thin enough to address pre-equilibrium interaction processes of ions in solids. The two distributions allow a separation of different impact parameter regimes. This implicates that the interaction of particles with ultra-thin solid targets may not be described in terms of an "average interaction per unit length".\\
Highly charged ions are produced in a room-temperature electron beam ion trap (EBIT) \cite{ZKO08} at the Ion Beam Center of the Helmholtz-Zentrum Dresden-Rossendorf. Xe ion charge states from $Q=10$ - 30 are selected utilizing an analyzing magnet. To prevent charge exchange processes within the beam-line or the experimental chamber the base pressure is kept below $5\cdot 10^{-9}$\,mbar for all experiments. The kinetic ion energy is adjusted by means of an electrostatic deceleration system in the range of 40 - 135\,keV (310 - 1050\,eV/amu). Free-standing carbon nanomembranes with a thickness of 1\,nm \cite{TBN09,TG12} from \textsc{CNM Technologies} Bielefeld, Germany, on a standard transmission electron microscopy (TEM) grid with an underlying lacey carbon support are mounted within the experimental chamber. The membranes consist of a self-assembled monolayer of 1-1'-biphenyl-4-thiol (H-(C$_6$H$_4$)$_2$-SH), which has been cross-linked (polymerized) and hydrogen depleted by low-energy electron irradiation \cite{TBN09}. Contaminations of the CNM with light elements (O, F, I) have been reported \cite{TBN09} and additional sulfur contaminations are observed using Auger-Electron-Spectroscopy. The obtained concentrations are well below 1\,at\% and are therefore neglected in the following discussion, i.e. the CNM is considered as pure carbon material. Possible hydrogen content of the CNM is assumed to be small due to hydrogen loss upon preparation of the CNM \cite{TG12}. A separate manipulation stage within the experimental chamber holds an electrostatic analyzer with a \textsc{Hamamatsu Photonics} channeltron for ion counting. This manipulation stage allows angle resolved transmission measurements with an acceptance angle of the analyzer of 1.6\,$^{\circ}$. The energy resolution of the analyzer is measured to be $\Delta E \approx 1.5\cdot 10^{-3} E$, giving reasonable accuracy for charge exchange measurements. The total uncertainty in energy loss determination ranges from 60\,eV to 200\,eV mainly due to limited measurement precision, i.e. number of counted ions. Note that the mean energy of the transmitted ions is deduced from the median of the distribution of the corresponding exit charge state. The electrostatic energy filter allows 5000\,V as maximum voltage, which leads to constraints in measurements of large charge exchanges. The primary ion beam is charge state analyzed without target by the electrostatic analyzer to check for charge exchange with residual gas atoms within the beam-line. For incident ion charge states above $Q=20$ ions are detected with a charge loss of $\Delta Q = Q_{in} - Q_{exit} =1$ to 3. However, the amount of ions with lower charge states than the primary one is 4 orders of magnitude smaller. Their contribution can therefore be neglected. \\
The intensity of the ion beam is kept below 5000\,ions/s with a typical beam diameter of 1.5\,mm, yielding an ion flux of about $10^{9}$\,cm$^{-2}$h$^{-1}$. Significant damage of the membrane might occur only for exposure times longer than 100\,h assuming a critical fluence of $10^{11}$\,cm$^{-2}$ where 1\,\% of the ions hit a previous damaged area of 10\,nm$^2$. No degradation of the CNM during transmission measurements is observed. \\
TEM and helium ion microscopy studies of the CNM reveal that no pores larger than 1\,nm in diameter exist. However, on a larger scale the membrane shows cracks ($10-50$\,$\mu$m) due to a non-perfect coverage over large holes in the support film. Control experiments using a TEM grid with a lacey carbon support film but without a CNM showed no charge exchange nor energy loss. Thus, we can conclude that (a) ions penetrating the lacey carbon film are either stopped within the film or transmitted as neutral atoms and are therefore not detected and (b) that the amount of ions undergoing small angle deflections on the walls of supporting structure or during passage through cracks is negligible. \\
For slow highly charged Xe ions transmitted through a CNM two distinct exit charge state distributions are observed. Fig. \ref{fig1} shows a typical transmission spectrum obtained with the electrostatic analyzer. The positions of the exit charge states are marked by arrows. The first distribution, ranging from $Q_{exit}=29$ to 12, shows an intensity maximum at $Q_{exit}=28$. Within the uncertainty of the measurement the peaks show no energy loss and no energy straggling. Note that the peak width originates from the channeltron entrance slit (detector resolution), whereas the uncertainty is determined by the steepness of the peak edges. The second distribution, ranging from $Q_{exit}=12$ to 5, is instead combined with an energy loss and an energy straggling visible as a larger peak width in Fig. \ref{fig1}. Due to the voltage limitation of our spectrometer the maximum of the second distribution can not be determined. \\
In order to distinguish between possible processes leading to the two exit charge state distributions angle resolved transmission measurements are performed. Fig. \ref{fig2} depicts three different transmission spectra for a 46.8\,keV Xe$^{25+}$ ion beam analyzed under 0\,$^{\circ}$, 2\,$^{\circ}$ and 4\,$^{\circ}$ projectile exit angle, respectively. Clearly the distribution of high charge states vanishes with tilting angle (see double logarithmic representation in the inset), while low charge states are transmitted up to 4\,$^{\circ}$, even though the intensity decreases more than one order of magnitude. The energy losses $\Delta E(\alpha _{exit})$ for ions with $Q_{exit}=2$ ($\Delta Q = 23$) are marked as well. From the increase in energy loss with deflection angle and the fact that the high charge state distribution is only observed in (exact) forward direction we conclude that the low charge state distribution results from close collisions of the ions with target atoms. Since the mass of Xe is approximately 10 times higher than the mass of C energy and momentum conservation yield a maximum deflection angle of 5.2\,$^{\circ}$ for one elastic scattering event. Due to the small thickness of 1\,nm of the membrane we expect that the ions undergo at most one scattering event \cite{footnote1}. In contrast, the high exit charge state distribution is only observed under straight forward direction and the energy loss is negligible. This is attributed to the fact that for large impact parameters the deflection angle as a result of an elastic collision as well as the transferred energy (energy loss) become very small. In this case a contribution to the energy loss could only result from ion interaction with the electrons of the membrane. However, the observed energy loss for the high exit charge states is in any case smaller than the measurement uncertainty. \\
The measured energy loss at 0\,$^{\circ}$ for highly charged ions with a constant kinetic energy of 40\,keV (310\,eV/amu) but varying incident charge states is shown in Fig. \ref{fig3}. Fig. \ref{fig3}(a) shows the mean energy loss as a function of the charge loss $\Delta Q$. The energy loss is strongly dependent on the charge loss, whereas the same charge loss leads to different energy losses depending on the residual charge (e.g. $\Delta Q = 18$ for $Q_{in}= 20$, 25 and 30, respectively (see Fig. \ref{fig3}(a))). The green dots in Fig \ref{fig3}(b) represent the energy loss deduced from the peaks with $\Delta Q =1$, i.e. from ions which exhibit the smallest charge loss as a function of the incident charge state $Q_{in}$. The energy loss for these ions lies within the measurement uncertainty and can only be estimated to be smaller than 60\,eV. The red dots in Fig. \ref{fig3}(b) show the energy loss obtained from the analysis of the $Q_{exit}=2$ peaks, i.e. for ions with the highest charge loss observable with our setup. These ions show a quadratic increase of the mean energy loss with incident charge state (see fit in Fig \ref{fig3}(b)) and consequently a much higher value than predicted by TRIM \cite{ZBZ08}. In fact, the TRIM result of 237\,eV reproduces the energy loss for a neutral atom ($Q_{in} =0$) from an extrapolation of our measured data (264\,eV) if one considers a CNM carbon density of $5.54\cdot 10^{22}$\,at/cm$^3$ \cite{MWK13} and an exit angle of $<1.6$\,$^{\circ}$ in the simulation. Since we identified low exit charge states resulting from nuclear (i.e. elastic) collisions, the charge state dependent energy loss represents experimental evidence for the predicted increase in nuclear stopping with projectile ion charge state by Biersack \cite{B93}, even though the predicted values are not reproduced. \\
To estimate the amount of ions transmitted at charge states, which can not be observed by the analyzer or even as neutrals, the transmission spectra have been normalized. For intensity normalization we employed the fact that a membrane usually does not cover the entire TEM grid perfectly but has some micrometer sized cracks. Ions which pass through these cracks do not interact with the target and therefore remain in their incident charge state without any energy loss. For a constant incident kinetic ion energy of 40\,keV the ratio of transmitted ions per exit charge state to the amount of transmitted ions through cracks is shown in Fig. \ref{fig5}. Note that all data shown in Fig. \ref{fig5} are obtained from the same sample. The two distributions can clearly be distinguished by the minimum in between (e.g. $Q_{exit} = 11$ for $Q_{in} =30$). Furthermore, the data indicates that for higher incident charge states more ions are neutralized. This can be derived from the steeper slopes towards lower exit charge states below the minimum for $Q_{in} = 30$ and 25 than for $Q_{in} = 20$, 15 and 10. This fact becomes more evident in Fig. \ref{fig6} where the integral of the curves from Fig. \ref{fig5} is plotted as a function of the incident charge state. An integrated normalized intensity of about 1 - 1.1 is obtained for incident charge states 10 to 20, whereas the value drops towards higher incident charge states to about 0.3. From the drop we conclude that for increasing incident charge states more ions are transmitted as single charged ions or as neutral atoms. Above a certain charge state the ion captures therefore electrons more effectively from the target system. The onset of the drop at about $Q_{in} = 20$ coincides very well with the observed potential energy threshold for pore production by HCI in CNM reported recently \cite{RWS13}. We conclude that the potential energy deposition becomes more efficient above a threshold charge state. \\
The fact that the two distributions are well separated may result from a strongly impact parameter dependent charge exchange. For close collisions, where also nuclear energy transfer occurs, the ion and the target atom (or target molecule) form a quasi-molecule due to the strong overlap of their corresponding electronic wave functions. This overlap leads to a strong level shift \cite{AAE97} and therefore a direct capture of target electrons into inner shells of the ion. The consideration of the biphenyl molecule (or rather the aromatic ring after cross-linking) as the "target molecule" for the charge exchange process is physically justified by the delocalization of the carbon valence electrons over the aromatic ring and to some extend over its neighboring molecules. It also provides enough electrons to reach values of up to $\Delta Q=28$. Contributions to the charge exchange from other species than carbon are neglected due to their small concentrations ($<1$\,at\%) and from hydrogen due to the fact that it provides only one electron per atom. For large impact parameters ($>2$\,\r{A}) the nuclear charge of the carbon atoms is sufficiently screened and no nuclear energy transfer occurs. Electrons may only be transferred to the ion via classical over barrier (COB) transport \cite{N86,BLM91}, because no overlap of the electron densities of the molecule and the ion occurs. The critical distance $R_c \propto \sqrt{8Q+2}/W$ for classical charge transfer depends on the work function $W$ of the material in the case of above surface neutralization \cite{BLM91}. The present experimental findings indicate a critical distance $R_c$ smaller than the interatomic distances in the membrane. Therefore the work function should be identified here with the ionization energy of the cross-linked biphenyl molecule or for simplicity with the atomic ionization energies (11\,eV, 24\,eV, 46\,eV, ...) for successive ionization of carbon \cite{T94}. Since the second electron to be transferred has already a much higher binding energy than the first one, its exchange is only possible at much shorter distances. Niehaus showed in his extended classical over barrier model for charge exchange of highly charged ions interacting with molecules that the cross section for three-electron-capture is already about a factor of 30 smaller than the cross section for one-electron-capture \cite{N86}. Charge exchanges for impact parameter larger than those needed for a sufficient level shift are therefore limited to small $\Delta Q$. For slow highly charged ions interacting with very thin target materials the processes can hence be described better in a picture of ion-molecule interaction rather than ion-solid interaction. The concept of stopping force as the \textit{mean} energy loss per unit length fails in a thin membrane, because no averaging over impact parameters appears anymore.
\begin{acknowledgments}
The authors acknowledge fruitful discussions with Christoph Lemell. Thanks go to Nico Klingner for support with the channeltron. Financial support from the Deutsche Forschungsgemeinschaft (DFG) (project-no. HE 6174/1-1) and from the Austrian FWF (project-no. I 1114-N20) is acknowledged.
\end{acknowledgments}

\bibliography{references}

\end{document}